\PassOptionsToPackage{dvipsnames}{xcolor}
\documentclass[10pt,twocolumn,letterpaper]{article}

\usepackage[pagenumbers]{wacv}               

\usepackage{multirow}
\usepackage[table]{xcolor}
\usepackage{accessibility}
\usepackage{amsmath}
\usepackage{graphicx}
\usepackage{booktabs}

\usepackage{caption}  

\definecolor{wacvblue}{rgb}{0.21,0.49,0.74}
\usepackage[pagebackref,breaklinks,colorlinks,allcolors=wacvblue]{hyperref}

\def\wacvPaperID{1}
\def\confName{WACV}
\def\confYear{2026}

\title{RoadBench: A Vision-Language Foundation Model and Benchmark for Road Damage Understanding}

\author{
\textbf{Xi Xiao}\textsuperscript{1,7,*}\quad
\textbf{Yunbei Zhang}\textsuperscript{2,*}\quad
\textbf{Janet Wang}\textsuperscript{2}\quad
\textbf{Lin Zhao}\textsuperscript{3}\quad
\textbf{Yuxiang Wei}\textsuperscript{4}\quad
\textbf{Hengjia Li}\textsuperscript{5}\\
\textbf{Yanshu Li}\textsuperscript{6}\quad
\textbf{Xinyuan Song}\textsuperscript{9}\quad
\textbf{Xiao Wang}\textsuperscript{7}\quad
\textbf{Swalpa Kumar Roy}\textsuperscript{8}\quad
\textbf{Hao Xu}\textsuperscript{10}\textsuperscript{\dag}\quad
\textbf{Tianyang Wang}\textsuperscript{1}\textsuperscript{\dag}\\[3mm]
\textsuperscript{1}University of Alabama at Birmingham, Birmingham, AL, USA\\
\textsuperscript{2}Tulane University, New Orleans, LA, USA\\
\textsuperscript{3}Northeastern University, Boston, MA, USA\\
\textsuperscript{4}Georgia Institute of Technology, Atlanta, GA, USA\\
\textsuperscript{5}Carnegie Mellon University, Pittsburgh, PA, USA\\
\textsuperscript{6}Brown University, Providence, RI, USA\\
\textsuperscript{7}Oak Ridge National Laboratory, Oak Ridge, TN, USA\\
\textsuperscript{8}Tezpur University, Assam, India\\
\textsuperscript{9}Emory University, Atlanta, GA, USA\\
\textsuperscript{10}Harvard University, Cambridge, MA, USA\\[2mm]
{\small \textsuperscript{*}Equal contribution.}\\
{\small \textsuperscript{\dag}Corresponding authors: \texttt{haxu@bwh.harvard.edu, tw2@uab.edu}}
}

\begin{document}

\maketitle

\begin{figure*}[t]
  \centering
  \vspace{-4mm}
  \includegraphics[width=\textwidth]{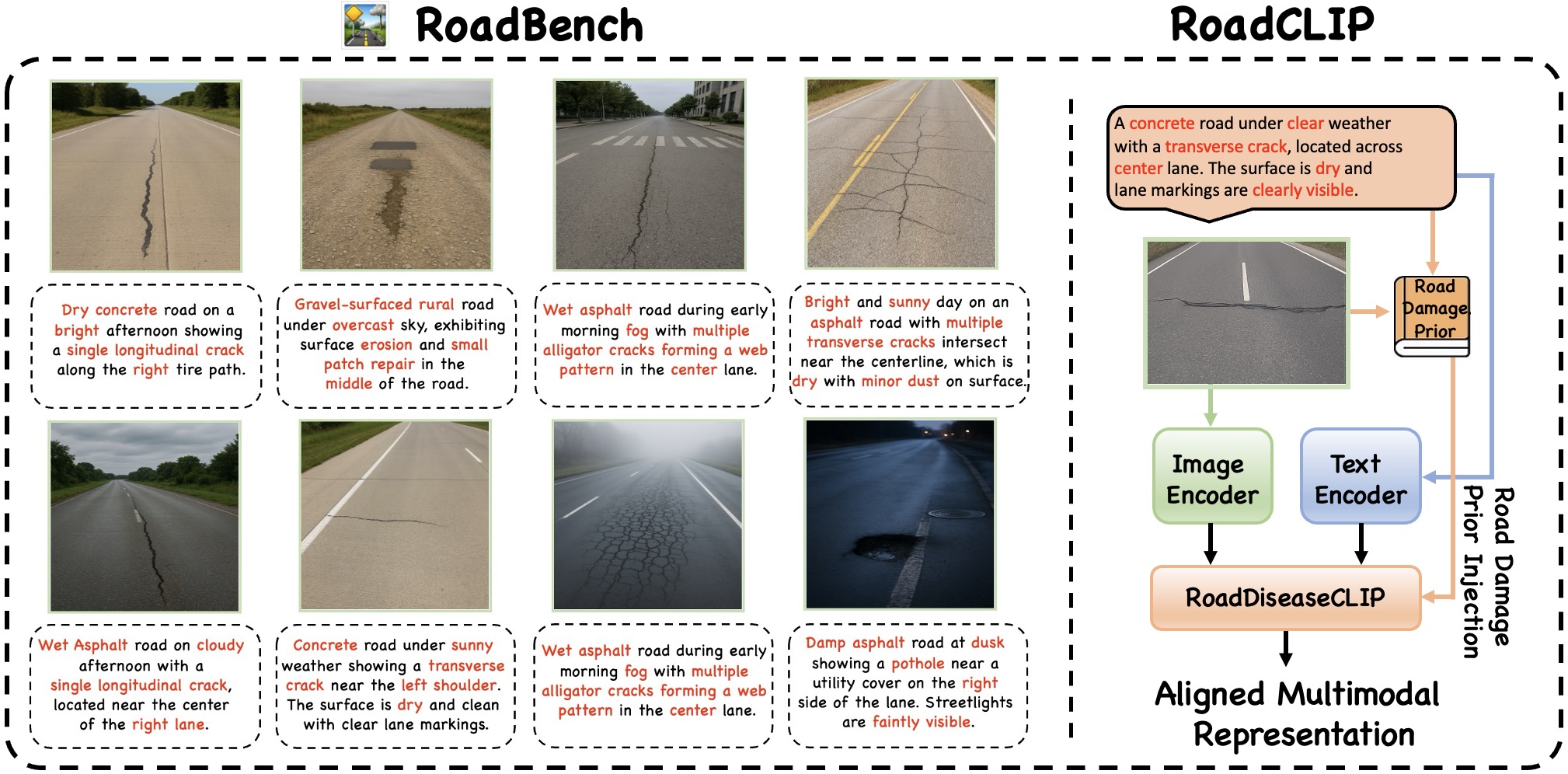}
  \vspace{-3mm}
  \caption{Overview of the \textbf{RoadBench} benchmark and the \textbf{RoadCLIP} framework.
    \textbf{Left:} Sample image–text pairs synthesized in diverse road scenarios, capturing damage types (e.g., longitudinal cracks, potholes), weather conditions, spatial context, and surface materials.
    \textbf{Right:} Our \textbf{RoadCLIP} architecture leverages a dual-encoder backbone enhanced with \textit{Disease-aware Positional Encoding} and \textit{Road Disease Prior Injection} to align visual and textual features in a multimodal embedding space.}
  \label{fig:ab_interpolation}
  \vspace{-3mm}
\end{figure*}

\begin{abstract}
Accurate road damage detection is crucial for timely infrastructure maintenance and public safety, but existing vision-only datasets and models lack the rich contextual understanding that textual information can provide. To address this limitation, we introduce \textbf{RoadBench}, the first multimodal benchmark for comprehensive road damage understanding. This dataset pairs high-resolution images of road damages with detailed textual descriptions, providing a richer context for model training. We also present \textbf{RoadCLIP}, a novel vision-language model that builds upon CLIP by integrating domain-specific enhancements. It includes a disease-aware positional encoding that captures spatial patterns of road defects and a mechanism for injecting road-condition priors to refine the model’s understanding of road damages. We further employ a GPT-driven data generation pipeline to expand the image–text pairs in \textbf{RoadBench}, greatly increasing data diversity without exhaustive manual annotation. Experiments demonstrate that \textbf{RoadCLIP} achieves state-of-the-art performance on road damage recognition tasks, significantly outperforming existing vision-only models by 19.2\%. These results highlight the advantages of integrating visual and textual information for enhanced road condition analysis, setting new benchmarks for the field and paving the way for more effective infrastructure monitoring through multimodal learning. 
\end{abstract}
\section{Introduction}
\label{sec:intro}

Road infrastructure is the backbone of economic development and social connectivity, supporting efficient transportation, commerce, and access to essential services. However, damages such as cracks, potholes, and pavement deformation significantly reduce ride quality and, if left unaddressed, pose serious safety risks, resulting in substantial economic costs. To tackle these challenges, the research community has turned to computer vision to develop automated systems for road damage detection~\cite{arya2021rdd2020, arya2024rdd2022, zhang2022new}. The emergence of several large-scale vision-only datasets (e.g., RDD2022~\cite{arya2024rdd2022} with over 47,000 images of road surfaces across six countries) has catalyzed the development of deep learning models for pavement distress identification and classification~\cite{bochkovskiy2020yolov4, ren2015faster, redmon2016you}.

However, existing models are exclusively vision-based, relying solely on visual cues (e.g., image features), without considering  textual descriptions of the damage~\cite{arya2021rdd2020, arya2024rdd2022}. As a result, these models often struggle to capture fine-grained distinctions, such as the severity of a crack or its precise location, and may fail to generalize across diverse environmental and road conditions.  Inspired by the remarkable success of multimodal approaches, like CLIP~\cite{radford2021learning}, which demonstrates a strong zero-shot recognition capability by learning joint image–text representations, we are motivated to explore whether multimodality could significantly improve accuracy, adaptability, and semantic understanding in real-world infrastructure monitoring scenarios. 
However, to our best knowledge, no publicly available dataset pairs road surface images with detailed textual descriptions, limiting the development of such multimodal solutions in this domain. Moreover, well-known large foundation models (e.g., GPT-4o~\cite{openai2023gpt4} and DeepSeek-VL~\cite{lu2024deepseekvl}) yield poor performance on road damage understanding (see Fig~\ref{fig:attention_vis} for details), a challenge often tied to the transferability of general-purpose representations to specialized tasks \cite{NEURIPS2024_d3602fc9, zhang2025dpcore, Zhang_2025_WACV}.

To fill this gap, we introduce \textbf{RoadBench}, the first multimodal benchmark for road damage understanding, along with \textbf{RoadCLIP}, the first vision–language model tailored for joint image–text learning in this domain, built on this dataset.
\textbf{RoadBench} includes 100,000 high-resolution road images, with each paired with a detailed and reliable description of the scene’s pavement condition, generated by a state-of-the-art generative language model (i.e., GPT-4o~\cite{openai2023gpt4}). The descriptions are employed to reflect real-world conditions, including diverse environments (e.g., urban vs. rural), varied weather and lighting, and a wide range of damage types and severities. We show several examples in Figure \ref{fig:ab_interpolation}.
\textbf{RoadCLIP} comprises two key modules: a \textit{disease-aware positional encoding} module and a \textit{domain-specific prior injection} mechanism. 
The \textit{disease-aware positional encoding} module injects knowledge of road geometry and common damage localization patterns into the visual branch of the multimodal backbone, making  the learned representations sensitive to the road surface, where a damage occurs (e.g., wheel-path cracks and shoulder cracks), and the spatial scope of the damage, well aligned with human inspectors' professional actions of checking both location and spread of a damage. 
The \textit{domain-specific prior injection} equips the model with expert knowledge about typical “road diseases” (e.g., the co-occurrence of certain crack patterns or textural cues indicating material fatigue). This prior is integrated during training for image–text alignment, acting as guidance that makes multimodal representations more discriminative for our task. With the two primary innovations, \textbf{RoadCLIP} learns a joint embedding space where images and descriptions of road damage are tightly aligned, facilitating more accurate cross-modal understanding. 

Then, we conduct extensive experiments to evaluate our dataset and model. Comparing with state-of-the-art baselines, \textbf{RoadCLIP} achieves superior results across multiple metrics. Notably, our approach outperforms the best purely visual model by \textbf{19.2\%} in detection accuracy and \textbf{20.9\%} in classification F\textsubscript{1}-score. It also surpasses a general vision–language baseline in image–text retrieval by a significant margin (e.g., Recall@1 improved by \textbf{14.9\%}). In addition, our ablation study reveals that either the positional encoding or the prior injection module plays a vital role in multimodal alignment and generalization of the proposed method.

Our primary contributions are summarized as follows: 
\begin{itemize}[left = 0em]
\item We establish \textbf{RoadBench}, the first image-and-text benchmark for road damage understanding, with a great potential of being used for training multimodal models and evaluating methods proposed in the realm. This dataset is also the largest one (i.e., with a size of 100,000) in this field, offering diverse samples for understanding road damages.   
\item We develop \textbf{RoadCLIP}, a new vision–language foundation model with a tailored architecture (including a novel positional encoding and prior knowledge injection) to effectively learn from road images and their descriptions. \textbf{RoadCLIP} achieves precise, fine-grained alignment between textual semantics and corresponding image regions (Figure~\ref{fig:text_region_alignment}), and demonstrates superior attention localization on road damage areas compared to GPT-4o and DeepSeek-VL (Figure~\ref{fig:attention_vis}). Moreover, it provides insights into designing domain-specific models in multimodal context.   
\item We demonstrate that our approach achieves state-of-the-art performance on road damage recognition tasks, significantly outperforming existing vision-only and multimodal methods.
\end{itemize} 
\section{Related Work}
\label{sec:related}
\begin{table*}[!ht]
    \centering
    \caption{Comparison of \textbf{RoadBench} with existing single-modal road damage datasets. Our dataset is the first to provide multimodal (vision+language) annotations at scale, supporting a wide range of tasks beyond traditional detection.}
    \label{tab:dataset_comparison}
    \rowcolors{2}{gray!5}{white}
    \resizebox{\textwidth}{!}{
    \begin{tabular}{
        >{\centering\arraybackslash}p{3.5cm}  
        >{\centering\arraybackslash}p{2.5cm}    
        >{\raggedleft\arraybackslash}p{2.2cm}    
        >{\centering\arraybackslash}p{2.5cm}    
        >{\centering\arraybackslash}p{4cm}    
        >{\centering\arraybackslash}p{4.5cm}  
        >{\centering\arraybackslash}p{2.2cm}    
    }
    \toprule
    \specialrule{0.8pt}{0pt}{0pt}
    \rowcolor[gray]{0.9}
    \textbf{Dataset} & \textbf{Modality} & \textbf{Samples} & \textbf{Resolution} & \textbf{Disease Types} & \textbf{Task Type} & \textbf{Availability} \\
    \specialrule{0.8pt}{0pt}{0pt}
    RDD2022~\cite{arya2024rdd2022} & Vision-only & 47,420 & Variable & Crack, Pothole & Detection & Public \\
    TD-RD~\cite{xiao2025tdrdtopdownbenchmarkrealtime} & Vision-only & 7,088 & 3840×2160 & Crack, Pothole, Repair & Detection & Public \\
    CRDDC'22~\cite{arya2022crowdsensing} & Vision-only & 11,720 & Variable & Crack, Patch, Spalling & Detection & Public \\
    CNRDD~\cite{zhang2022new} & Vision-only & 9,053 & Variable & Crack, Patch, Rutting & Detection & Public \\
    GAPs384~\cite{cha2024deep} & Vision-only & 1,969 & 1920×1080 & Crack, Pothole & Detection & Public \\
    CFD~\cite{shi2016automatic} & Vision-only & 1,180 & 480×320 & Crack & Detection & Public \\
    \midrule
    \rowcolor[gray]{0.95}
    \textbf{\textbf{RoadBench} (Ours)} & \textbf{Vision+Text (Multimodal)} & \textbf{100,000 } & \textbf{3840×2160} & \textbf{Ten types in all, more details see Fig\ref{fig:category_pie}} & \textbf{Multimodal Detection, Retrieval, Captioning, QA} & \textbf{Public} \\
    \bottomrule
    \end{tabular}
    }
\end{table*}

\subsection{Road Damage Detection Benchmarks} Existing road damage datasets and benchmarks are predominantly vision-only, focusing on detecting and localizing road surface defects from images~\cite{arya2022crowdsensing,arya2021rdd2020}. The Road Damage Dataset 2020 (RDD2020)~\cite{arya2021rdd2020} contains over 26,000 road images from multiple countries with more than 31,000 annotated damage instances, while the recent RDD2022 dataset~\cite{arya2024rdd2022} expanded this to 47,420 images across six countries, featuring over 55,000 labeled instances. Released as part of the CRDDC 2022 competition~\cite{arya2022crowdsensing}, these datasets have spurred the development of robust detection models using advanced architectures like YOLO series~\cite{redmon2016you,bochkovskiy2020yolov4,yu2021pp} and Transformer-based~\cite{carion2020end,zhu2021deformable,zhao2024detrs,fang2021you,li2023lite,zhang2021vit}. Other specialized datasets include TD-RD~\cite{xiao2025tdrd}, which offers aerial-view images from drone photography, GAPs384~\cite{cha2024deep} and CFD~\cite{shi2016automatic} for specific damage types, and CNRDD~\cite{zhang2022new} featuring Chinese road conditions. However, the critical limitation across all these datasets is their unimodal nature—they contain only images and annotations without any natural language descriptions, which restricts contextual understanding, generalization capabilities \cite{zhang2025dpcore}, and semantic reasoning about damage characteristics~\cite{sharma2018conceptual,subramanian2020medicat,wang2024ensemble,li2024yolox,dugalam2024development,zhang2024real,yu2024road,9904187,wang2024drivedreamer,wu2024seesr,he2024can,wu2024one}. Detailed information on these datasets is presented in Table \ref{tab:dataset_comparison}.

\subsection{Vision-Language Models} 
There have been rapid advances in vision-language models that jointly learn from images and text~\cite{li2025benchmark,li2022elevater,hong2020more,kuckreja2024geochat,jain2024semantic,dong2024changeclip,marino2019ok,li2024survey,chen2023benchmarking,roberts2024image2struct,chen2024viseval,zheng2022vlmbench,hong2024cogagent,wang2024cogvlm,yao2024tcp,huang2024adapting,li2024cycleyoloefficientrobustframework}. Contrastive models like CLIP~\cite{radford2021learning} align visual and textual representations using large-scale image–text pairs, enabling strong zero-shot performance. Unified architectures such as BLIP~\cite{li2022blip} and Florence~\cite{yuan2021florence} extend these capabilities to both understanding and generation through web-scale multimodal pretraining. Recent advances have further integrated powerful language models with visual processing~\cite{li2024seed,bitton2023visit,liu2024ii,lin2024vila,hong2024cogagent,kayser2021vil,tankala2024wikido,wang2024journeybench,yin2023lamm,zhang2024unveiling}, including techniques like test-time adaptation \cite{Zhang_2025_WACV} and prompt tuning \cite{xiao2025visualinstanceawareprompttuning, xiao2025visualvariationalautoencoderprompt}, as seen in models like LLaVA~\cite{liu2023llava}, GPT-4V~\cite{openai2023gpt4}, and Flamingo~\cite{alayrac2022flamingo}, which support sophisticated multimodal reasoning and contextual understanding. 
Evaluation typically uses benchmarks such as MS COCO Captions~\cite{lin2014microsoft} and Flickr30k~\cite{plummer2015flickr30k} for captioning and retrieval, VQAv2~\cite{goyal2017making} for visual question answering, RefCOCO~\cite{yu2016modeling} for referring expressions, and ScienceQA~\cite{lu2022learn} for scientific reasoning. Despite this progress, a significant gap exists in domain-specific multimodal benchmarks. While similar efforts are emerging in other fields like medicine \cite{xiao2025medicalimages}, none of the existing evaluation frameworks target specialized applications like road damage detection or infrastructure assessment, limiting our understanding of how these powerful models perform in critical real-world domains where both visual recognition and linguistic description are essential. .

\section{Dataset Construction}
\label{sec:dataset_construction}
In this section, we present \textbf{RoadBench}, the first global multimodal benchmark dataset designed for road damage detection, featuring paired high-resolution road images and corresponding detailed textual descriptions.

\subsection{Data Curation}

\begin{figure}[!ht]
    \centering
    \includegraphics[width=0.95\linewidth]{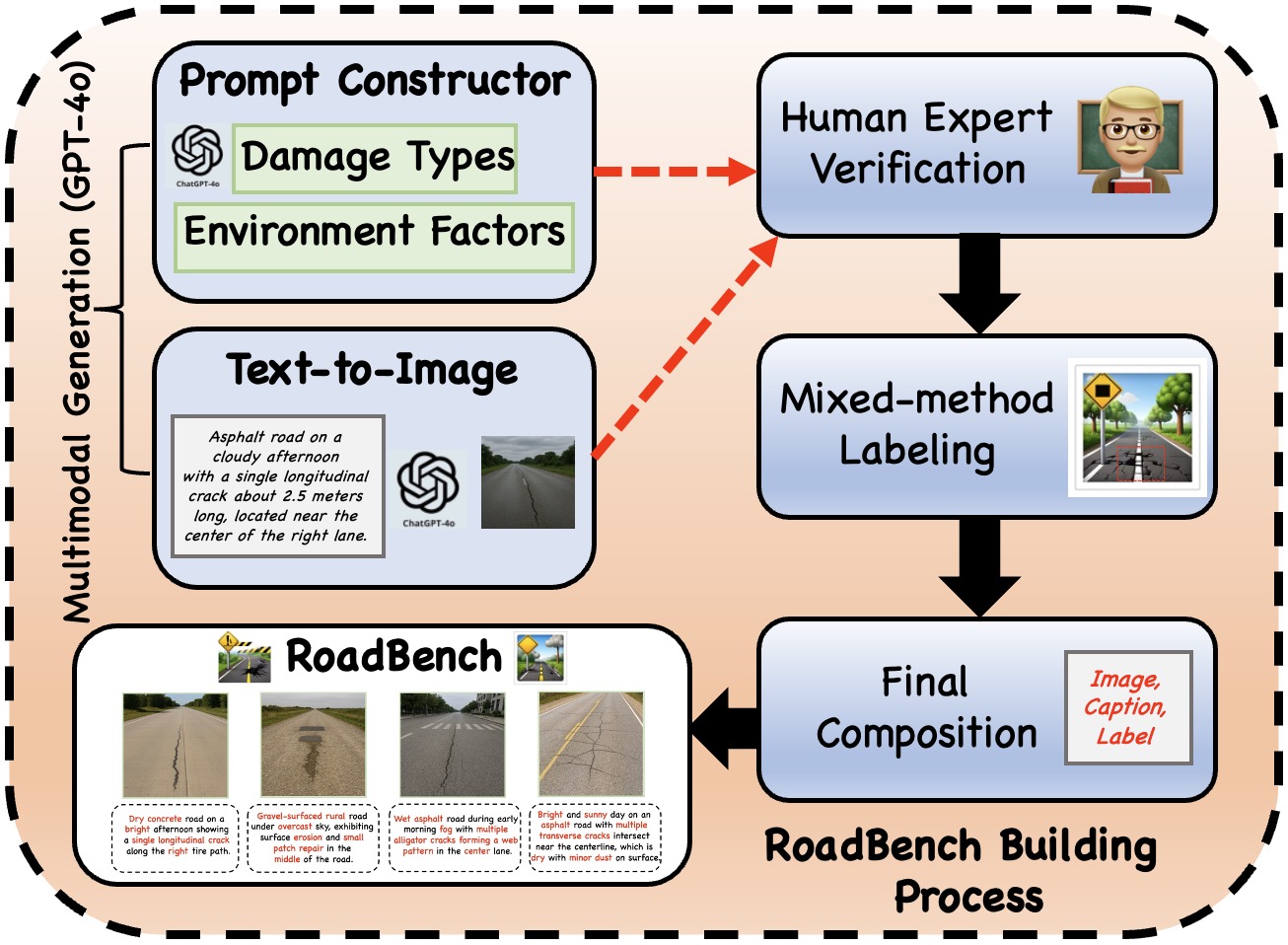}
    \caption{Overview of the \textbf{RoadBench} construction pipeline. Structured prompts describing road damage types and environments guide multimodal generation with GPT-4o. Human experts verify the generated image–text pairs, which are then annotated and compiled into a high-quality benchmark dataset with images, captions, and labels.}
    \label{fig:roadbench_pipeline}
\end{figure}

\textbf{Data Generation.}  
We firstly consult domain experts to establish a standardized vocabulary including road disease types, environmental conditions, and beyond. This expert-informed terminology guides the construction of textual prompts, which are subsequently utilized by the multimodal GPT-4o model to synthesize high-resolution (3840×2160 pixels) road surface images accompanied by corresponding textual descriptions. The generated outputs are subsequently subjected to human expert verification to ensure accuracy and relevance, a practice also found effective in other domains like medical image synthesis \cite{wang2025doctor}. A mixed-method labeling approach is then applied to enhance label quality, followed by a final composition step that integrates the image, caption, and corresponding label into the \textbf{RoadBench} dataset.

\noindent
\textbf{Data Annotation.} 
For each road damage image, we firstly annotate the location of the damage using manual labeling or using a generative approach and then generate the corresponding binary masks. These spatial annotations complement the textual descriptions, providing valuable priors for multimodal models and enabling tasks such as text-guided damage localization.

\noindent
\textbf{Data Validation.}
Civil engineering experts reviewed each image-text pair to validate the visual realism of the images and the accuracy of the corresponding textual descriptions. An iterative refinement process was also applied—flagged samples were systematically regenerated or edited to support consistent, high-quality standards across the dataset.

\subsection{Data Composition}
\textbf{RoadBench} contains 100{,}000 image–text pairs, each carefully generated to represent a wide range of realistic road damage scenarios. Every textual label provides precise details, including damage dimensions (e.g., a 2-meter crack), spatial positioning (e.g., center or shoulder), and environmental context (e.g., bright, wet, foggy).
\begin{figure}[!ht] 
\centering 
\includegraphics[width=0.85\linewidth]{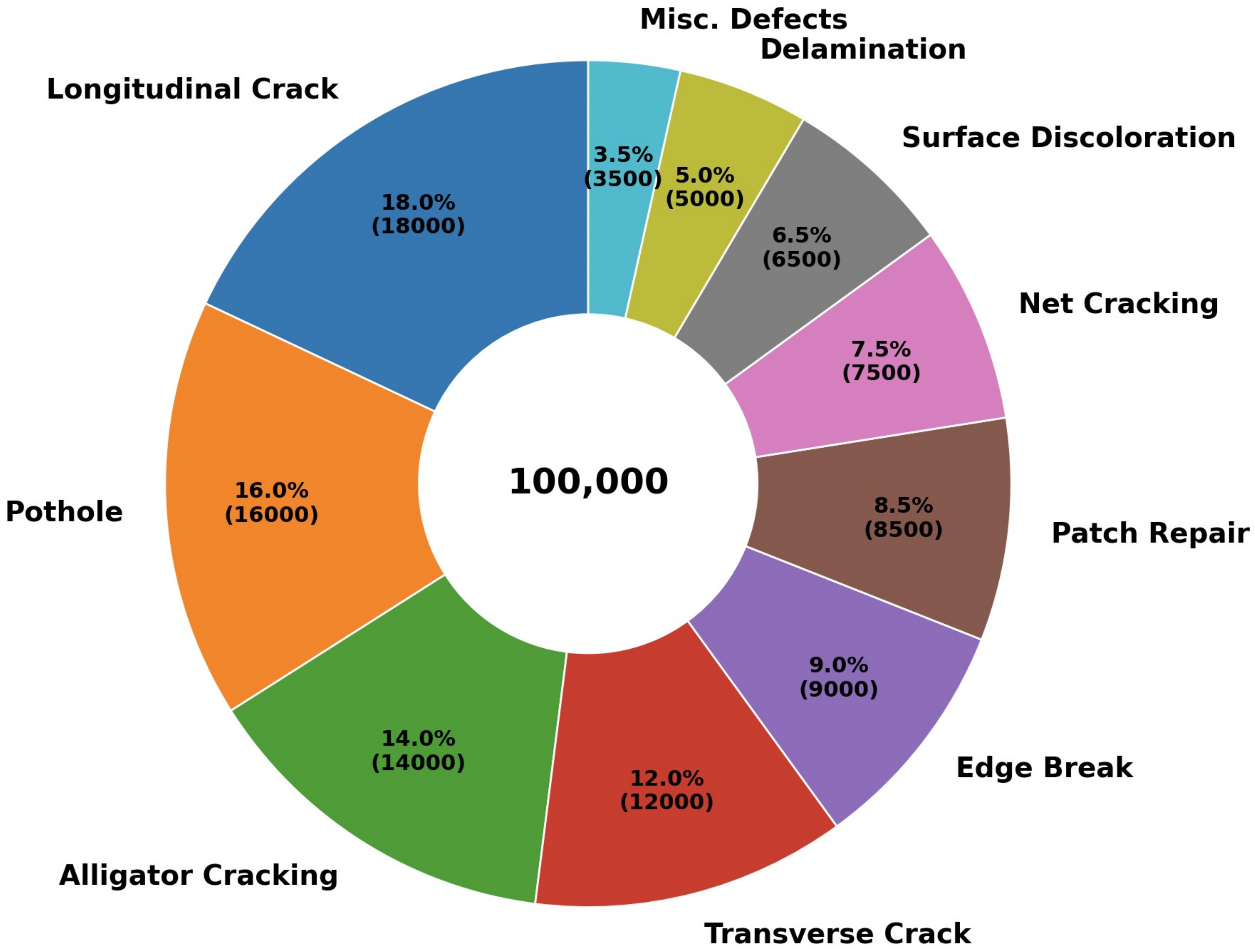} \caption{Category-wise proportion of road defect types in \textbf{RoadBench}.} \label{fig:category_pie} 
\end{figure}
The dataset cover 10 major types of road surface defects — \textit{longitudinal cracks}, \textit{transverse cracks}, \textit{alligator cracking}, \textit{potholes}, \textit{patch repair}, \textit{edge cracks}, \textit{centerline cracks}, \textit{discoloration}, \textit{mixed damage patterns}, and \textit{irregular/unknown defects}. 
As shown in Figure \ref{fig:category_pie}, \textbf{RoadBench} features a reasonably balanced class distribution, with deliberate inclusion of rare yet critical defect types to simulate real-world inspection conditions and support generalized model performance. Detailed statistics of the dataset can be found in supplementary materials.

\subsection{Is Synthetic Data Reliable?}

Recent advances in data-driven models have highlighted the importance of large, diverse, and high-quality datasets. However, collecting and annotating real-world data is often costly, time-consuming, and privacy-sensitive, particularly in domains such as autonomous driving or road damage assessment. Consequently, synthetic data generation has emerged as a viable alternative \cite{wang2025doctor, chung2025soksyntheticimagesreplace}, enabling controlled sampling, scalable labeling, and simulation of rare or hazardous scenarios. The reliability of synthetic data has been studied across various vision tasks. For example, Richter et al.~\cite{richter2016playing} demonstrated that synthetic datasets generated from video game engines (e.g., GTA V) can effectively train semantic segmentation models. Similarly, Tremblay et al.~\cite{tremblay2018training} showed that models trained on synthetic objects can generalize well to real-world object detection. More recently, synthetic datasets like Synscapes~\cite{wrenninge2018synscapes} and CARLA~\cite{dosovitskiy2017carla} have been widely adopted in autonomous driving research due to their realism and label precision. In this work, synthetic data serves a complementary role to real-world annotations. By leveraging a simulation pipeline augmented with text-guided prompt generation and selective augmentation, we ensure that the synthesized samples retain semantic fidelity and style diversity while remaining free of privacy concerns such as faces or license plates.

\section{Methodology}

\begin{figure*}[!ht]
  \centering
  \includegraphics[width=0.95\textwidth]{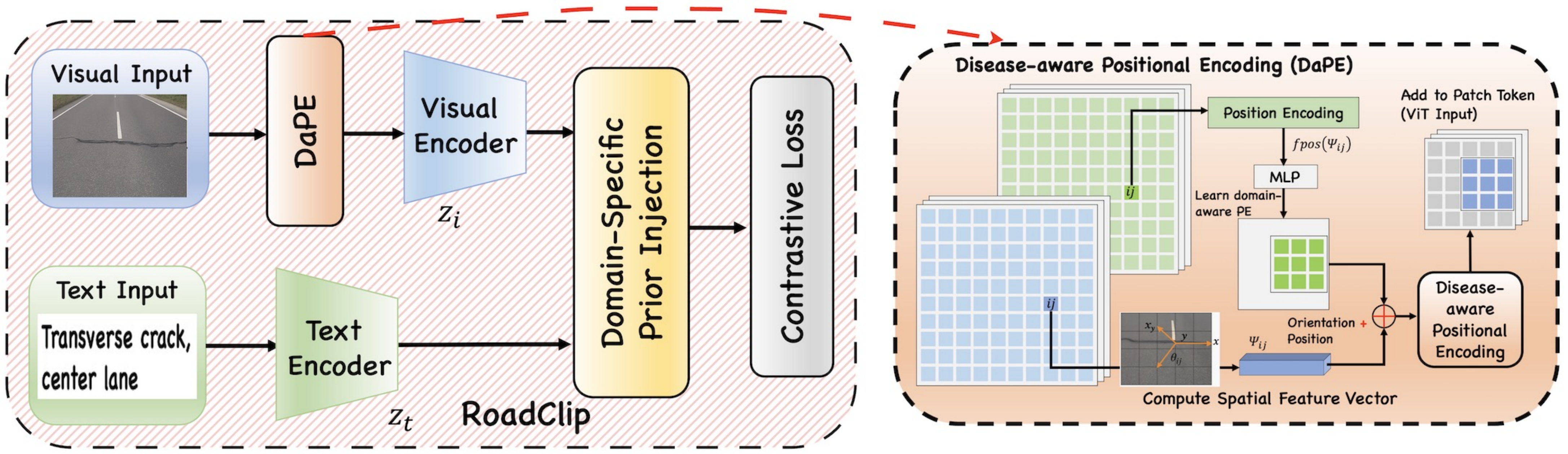}
  \caption{Overall architecture of \textbf{RoadCLIP}. The model uses a dual-encoder CLIP-based architecture, projecting road images and damage descriptions into a shared space, trained using a symmetric contrastive loss. A Disease-aware Positional Encoding (DaPE) module adds spatial priors to the visual encoder, while a Domain-Specific Prior Injection module enriches both modalities.}
  \label{fig:roadclip_architecture}
\end{figure*}

\subsection{Overall Architecture} 
\textbf{RoadCLIP} adopts a dual-encoder architecture building upon the CLIP framework~\cite{radford2021learning}, as shown in Figure~\ref{fig:roadclip_architecture}. It consists of a vision encoder $f(\cdot)$ and a text encoder $g(\cdot)$ that project images and textual descriptions, respectively, into a shared $d$-dimensional latent space.  
In the vision module, we use a Transformer-based image encoder (as in CLIP, e.g. a ViT) that processes the input image as a sequence of patch embeddings. The text module is a Transformer language model encoding the input description or label. Both encoders produce a feature vector of dimension $d$ (using the Transformer’s \textit{[CLS]} token or global pooled representation), which is then $\ell_2$-normalized and passed through a linear projection (also of dimension $d$) to produce the final image embedding $\mathbf{z}_i$ and text embedding $\mathbf{z}_t$. 
To adapt \textbf{RoadCLIP} for \emph{road-damage-aware} representation learning,  we incorporate two novel modules into its architecture: a Disease-aware Positional Encoding (DaPE) module integrated into the vision encoder, and a Domain-Specific Prior Injection mechanism affecting both the image and text representations.

\subsection{Disease-aware Positional Encoding (DaPE)}
\label{sec:dape}

While standard vision Transformers typically use generic positional encodings—either fixed sine-cosine patterns or learned absolute embeddings—such representations are often agnostic to the domain-specific spatial semantics needed in road damage analysis. In road scenes, spatial attributes such as crack orientation, position relative to lane markers, and edge proximity are critical for accurate classification and grounding. To this end, we propose a \textit{Disease-aware Positional Encoding (DaPE)} module that encodes both absolute and geometric priors tied to road “disease” patterns.

\vspace{4pt}
\noindent\textbf{Geometric-aware Spatial Embedding.}
For each image patch at normalized coordinates $(x, y) \in [0,1]^2$, we define a domain-specific spatial descriptor:
\begin{equation}
    \Psi_{ij} = [x, y, \cos\theta_{ij}, \sin\theta_{ij}],
\end{equation}
where $\theta_{ij}$ represents the dominant orientation angle of any crack-like structure in patch $(i,j)$. This angle is computed using texture analysis techniques (e.g., Sobel gradients or structure tensor). The descriptor $\Psi_{ij}$ thus encodes both absolute position and directional features.

\vspace{4pt}
\noindent\textbf{MLP-based Positional Encoding.}
This spatial tuple is passed through a learnable MLP $f_{\mathrm{pos}}$ to yield a $d$-dimensional positional vector:
\begin{equation}
    \mathbf{p}_{ij}^{\mathrm{(DaPE)}} = f_{\mathrm{pos}}(\Psi_{ij}) \in \mathbb{R}^d.
\end{equation}
This vector is added to the patch’s visual feature embedding, either at the input stage or as a positional bias in subsequent attention layers. As a result, spatially meaningful cues—such as being near the edge or aligned along a directional crack—are encoded directly into the model’s representation.

\vspace{4pt}
\noindent\textbf{Domain-aware Modeling.}
By replacing the standard positional encodings with $\mathbf{p}_{ij}^{\mathrm{(DaPE)}}$, we guide the model’s self-attention to consider spatial structures that are diagnostically meaningful. For instance, patches along a continuous horizontal crack will have similar orientation codes, encouraging them to cluster in feature space. Similarly, patches near the road edge may carry different priors than center-lane patches.

\subsection{Domain-Specific Prior Injection} 
General-purpose vision-language models like CLIP, trained on open-world data, often lack fine-grained understanding of domain-specific semantics. To bridge this gap, \textbf{RoadCLIP} incorporates structured domain knowledge—specifically, road damage categories—into the training process via \emph{Domain-Specific Prior Injection}, aligning visual and textual representations with semantic prototypes such as \textit{pothole}, \textit{longitudinal crack}, and \textit{patch repair}.

\noindent
\textbf{Concept Embedding Initialization.}  
We define a set of $K$ road damage types $\mathcal{C} = \{c_1, c_2, \dots, c_K\}$. Each class $c_k$ is associated with a joint image-text concept embedding, initialized in two ways: (1) \textit{Text-based}, where a descriptive phrase (e.g., ``a photo of a $c_k$ on a road'') is encoded by the text encoder $g(\cdot)$ to obtain $\mathbf{t}_{c_k}$; or (2) \textit{Learnable prototype}, a trainable vector $\mathbf{v}_{k} \in \mathbb{R}^d$ initialized from $\mathbf{t}_{c_k}$. We adopt the latter, enabling the model to refine semantic priors during training while preserving interpretability.

\noindent
\textbf{Prior-Aligned Training Objective.}  
For each image $I_i$ with damage category $y_i \in \mathcal{C}$, we align its projected feature $\mathbf{z}_i = f(I_i)$ to its corresponding concept embedding. We define the alignment loss:
\begin{equation}
L_{\mathrm{concept}}(I_i) = -\log \frac{\exp(\mathrm{sim}(\mathbf{z}_i, \mathbf{t}_{y_i})/\tau)}{\sum_{c \in \mathcal{C}} \exp(\mathrm{sim}(\mathbf{z}_i, \mathbf{t}_{c})/\tau)},
\end{equation}
where $\mathrm{sim}(\cdot,\cdot)$ is cosine similarity and $\tau$ is a temperature parameter. This cross-entropy loss encourages image features to cluster around their concept prototypes in the embedding space.

\noindent
\textbf{Joint Training with Contrastive Learning.}  
The concept alignment loss is integrated with CLIP’s standard image-text contrastive objective. Additionally, descriptive sentences for each damage type are periodically encoded to further anchor $\mathbf{v}_k$ in natural language space, enhancing textual generalization.

\subsection{Training Objective}

\textbf{RoadCLIP} is trained end-to-end with a composite loss that integrates contrastive alignment, concept-level supervision, and spatial robustness regularization.

\noindent
\textbf{Image-Text Contrastive Loss.}
The primary objective is a bidirectional InfoNCE loss~\cite{oord2018representation} encouraging high similarity between matched image–text pairs $(I_i, T_i)$ and low similarity for mismatches. Let $\mathbf{z}_i = f(I_i)$ and $\mathbf{z}_i^+ = g(T_i)$ denote the embeddings from the image and text encoders, respectively. The loss is:

\begin{multline}
\mathcal{L}_{\mathrm{ITC}}
  = -\frac{1}{N} \sum_{i=1}^{N}
      \Bigl[
         \log \frac{\exp\bigl(\mathrm{sim}(\mathbf{z}_i,\mathbf{z}_i^+)/\tau\bigr)}
                  {\sum_{j=1}^N \exp\bigl(\mathrm{sim}(\mathbf{z}_i,\mathbf{z}_j^+)/\tau\bigr)}
       \\
         +\;
         \log \frac{\exp\bigl(\mathrm{sim}(\mathbf{z}_i^+,\mathbf{z}_i)/\tau\bigr)}
                  {\sum_{j=1}^N \exp\bigl(\mathrm{sim}(\mathbf{z}_j^+,\mathbf{z}_i)/\tau\bigr)}
      \Bigr]
\end{multline}

\noindent
\textbf{Concept Prior Loss.}
To reinforce category-specific alignment, we apply a concept supervision loss $\mathcal{L}_{\mathrm{concept}}(I_i)$ based on known damage type labels $y_i$ for all images. This loss aligns image embeddings with learnable category prototypes $\mathbf{v}_{y_i}$ and is weighted by $\lambda_{\mathrm{concept}}$:

\begin{equation}
\mathcal{L}_{\mathrm{domain-align}} = \frac{1}{N} \sum_{i=1}^{N} \mathcal{L}_{\mathrm{concept}}(I_i)
\end{equation}

\noindent
\textbf{Position Consistency Loss.}
To regularize DaPE, we introduce $\mathcal{L}_{\mathrm{pos-consist}}$, encouraging stability under small spatial shifts (translation/rotation) by minimizing the discrepancy between original and perturbed embeddings:

\begin{equation}
\mathcal{L}_{\mathrm{pos-consist}}(I_i) = \left\| f(I_i) - f(I_i') \right\|_2^2
\end{equation}

\noindent
\textbf{Total Loss.}
The final training loss is:

\begin{equation}
\mathcal{L}_{\mathrm{total}} = \mathcal{L}_{\mathrm{ITC}} + \lambda_{\mathrm{concept}} \mathcal{L}_{\mathrm{domain-align}} + \lambda_{\mathrm{pos}} \mathcal{L}_{\mathrm{pos-consist}}
\end{equation}

We optimize all model parameters (vision/text encoders, DaPE, projection layers, and concept prototypes) using Adam, with temperature $\tau$ learnable. This multi-objective training yields spatially aware and semantically aligned representations, enabling accurate road damage recognition and cross-modal retrieval.

\begin{table*}[!ht]
    \centering
    \caption{Comprehensive performance comparison of RoadCLIP with single-modal and multimodal state-of-the-art methods on the RoadBench dataset. Best results are highlighted in \textbf{bold}, and second-best results are underlined.}

    \resizebox{\textwidth}{!}{
    \begin{tabular}{lcc|lcccccc}
        \toprule
        \specialrule{0.8pt}{0pt}{0pt}
        \rowcolor[gray]{0.9}
        \multicolumn{3}{c|}{\textbf{Single-modal Vision-only Methods}} & \multicolumn{7}{c}{\textbf{Multimodal Vision-Language Models}} \\
        \rowcolor[gray]{0.9}
        \textbf{Method} & Modality & SLA(\%) & \textbf{Method} & Modality & ZS Acc.(\%) & Recall@1(\%) & Recall@5(\%) & Recall@10(\%) & SLA(\%) \\
        \specialrule{0.8pt}{0pt}{0pt} 
        \arrayrulecolor{black} 
        YOLOv10-n \textcolor{lightgray}{\scriptsize{[NeurIPS24]}}~\cite{wang2024yolov10} & Vision & 46.5 & CLIP \textcolor{lightgray}{\scriptsize{[ICML21]}}~\cite{radford2021learning} & Vision+Text & 63.8 & 40.3 & 63.2 & 71.4 & 41.5 \\
        YOLOv10-s \textcolor{lightgray}{\scriptsize{[NeurIPS24]}}~\cite{wang2024yolov10} & Vision & 47.3 & BLIP-2 \textcolor{lightgray}{\scriptsize{[ICML23]}}~\cite{li2023blip} & Vision+Text & 67.9 & 43.7 & 66.5 & 74.0 & 46.3 \\
        YOLOS-ti \textcolor{lightgray}{\scriptsize{[NeurIPS21]}}~\cite{Fang2021YouOL} & Vision & 45.3 & LLaVA \textcolor{lightgray}{\scriptsize{[NeurIPS23]}}~\cite{liu2023llava} & Vision+Text & 69.1 & 46.0 & 68.9 & 76.2 & 48.0 \\
        YOLOS-s \textcolor{lightgray}{\scriptsize{[NeurIPS21]}}~\cite{Fang2021YouOL} & Vision & 46.8 & MiniGPT-4 \textcolor{lightgray}{\scriptsize{[arXiv23]}}~\cite{zhu2023minigpt} & Vision+Text & 68.7 & 44.9 & 67.5 & 75.3 & 47.1 \\
        PP-PicoDet \textcolor{lightgray}{\scriptsize{[arXiv21]}}~\cite{yu2021pppicodetbetterrealtimeobject} & Vision & 47.0 & InstructBLIP \textcolor{lightgray}{\scriptsize{[NeurIPS23]}}~\cite{dai2023instructblip} & Vision+Text & 70.2 & 47.6 & 70.1 & 77.8 & 49.2 \\
        RT-DETR \textcolor{lightgray}{\scriptsize{[CVPR24]}}~\cite{Zhao_2024_CVPR} & Vision & \underline{48.6} & mPLUG-Owl \textcolor{lightgray}{\scriptsize{[arXiv23]}}~\cite{ye2023mplug} & Vision+Text & 71.0 & 49.3 & 71.8 & 78.6 & 50.7 \\
        Lite-DETR \textcolor{lightgray}{\scriptsize{[CVPR23]}}~\cite{li2023lite} & Vision & 45.3 & Otter \textcolor{lightgray}{\scriptsize{[arXiv23]}}~\cite{li2023ottermultimodalmodelincontext} & Vision+Text & 71.4 & 49.7 & 72.2 & 79.1 & 50.9 \\
        FR-CNN \textcolor{lightgray}{\scriptsize{[NeurIPS15]}}~\cite{ren2015faster} & Vision & 39.9 & GPT-4V \textcolor{lightgray}{\scriptsize{[arXiv23]}}~\cite{openai2023gpt4} & Vision+Text & \underline{74.5} & \underline{52.6} & \underline{74.0} & \underline{80.8} & \underline{55.4} \\
        SSD-VGG16 \textcolor{lightgray}{\scriptsize{[ECCV16]}}~\cite{liu2016ssd} & Vision & 38.7 & DeepSeek-VL \textcolor{lightgray}{\scriptsize{[arXiv24]}}~\cite{lu2024deepseekvl} & Vision+Text & 72.9 & 50.1 & 72.3 & 79.5 & 53.0 \\
        TD-YOLOV10 \textcolor{lightgray}{\scriptsize{[ICASSP25]}}~\cite{xiao2025tdrdtopdownbenchmarkrealtime} & Vision & \textbf{49.0} & \textbf{\textbf{RoadCLIP} (Ours)} & Vision+Text & \textbf{78.6} & \textbf{58.4} & \textbf{78.1} & \textbf{84.6} & \textbf{61.9} \\
        \bottomrule
    \end{tabular}}
    \label{tab:comprehensive_comparison}
\end{table*}

\section{Experiments}
In this section, we present comprehensive experiments to validate the effectiveness of our proposed \textbf{RoadCLIP} model and the multimodal benchmark \textbf{RoadBench} dataset. 
\subsection{Experimental Setup}

\textbf{Datasets.}
We evaluate model performance primarily on our \textbf{RoadBench} dataset and further evaluate cross-dataset generalization on three widely used visual-only datasets: TD-RD~\cite{xiao2025tdrd}, {CNRDD}~\cite{zhang2022new}, and {CRDDC'22}~\cite{arya2022crowdsensing}.

\noindent
\textbf{Evaluation Metrics.}
\textit{Zero-Shot Classification Accuracy (ZS Acc.)} measures the model’s ability to classify road damage categories without seeing labeled training data for those specific types, reflecting semantic alignment and domain adaptation capability. \textit{Image-Text Retrieval Accuracy} is reported as Recall@$k$ (with $k=1,5,10$), evaluating how accurately the model retrieves the correct caption given an image (or vice versa), a standard protocol in vision-language literature~\cite{radford2021learning,li2023blip}. Additionally, we introduce \textit{Semantic Localization Accuracy (SLA)} to quantify the model’s ability to correctly ground and localize defects in the image guided by textual descriptions—a task inspired by recent advances in referring expression grounding~\cite{yang2023semantics}. SLA is computed by measuring IoU overlap between predicted and annotated binary masks generated via text-guided attention.

\noindent
\textbf{Implementation Details.}
We firstly initialize our model with CLIP~\cite{radford2021learning} pretraining weights and then fine-tune on \textbf{RoadBench} using a contrastive InfoNCE-style objective. Unless otherwise stated, the batch size is set to 128, learning rate to $1e^{-4}$, and training runs for 20 epochs. 

\subsection{Results}

We compare \textbf{RoadCLIP} with a range of vision-only and multimodal models. As shown in Table~\ref{tab:comprehensive_comparison}, vision-only detectors such as YOLOv10~\cite{wang2024yolov10}, PP-PicoDet~\cite{yu2021pppicodetbetterrealtimeobject}, and RT-DETR~\cite{Zhao_2024_CVPR} achieve moderate performance in Semantic Localization Accuracy (SLA), with TD-YOLOv10 reaching the highest score of 49.0\%. However, these models lack the ability to leverage language supervision, limiting their performance in retrieval and zero-shot classification tasks. Multimodal vision-language models show stronger results across all metrics. GPT-4V~\cite{openai2023gpt4} and DeepSeek-VL~\cite{lu2024deepseekvl} achieve 74.5\% and 72.9\% in zero-shot accuracy (ZS Acc.), respectively, and perform competitively in image-text retrieval. In contrast, \textbf{RoadCLIP} achieves consistent improvements across all evaluation tasks. It obtains \textbf{78.6\%} in ZS Acc., outperforming GPT-4V and DeepSeek-VL by \textbf{+4.1\%} and \textbf{+5.7\%}, respectively. On retrieval, \textbf{RoadCLIP} achieves \textbf{58.4\%} Recall@1 and \textbf{84.6\%} Recall@10, establishing new state-of-the-art performance on \textbf{RoadBench}. In SLA, \textbf{RoadCLIP} surpasses GPT-4V by \textbf{+6.5\%} and DeepSeek-VL by \textbf{+8.9\%}, highlighting its advantage in spatial grounding. These results verify the effectiveness of our proposed Disease-aware Positional Encoding (DaPE) and domain-specific prior injection strategy. The improvements are achieved without relying on large-scale general-domain pretraining, but rather through targeted multimodal alignment tailored for the road damage domain. In summary, \textbf{RoadCLIP} provides a strong and reliable foundation for multimodal understanding in structured environments. We similarly provide comparative experiments with purely visual metrics, see supplementary materials for more details. 

\subsection{Evaluating the Effectiveness of DaPE}
We evaluate the effectiveness of our Disease-aware Positional Encoding (DaPE) through an ablation study against sinusoidal, learnable absolute, relative, and no positional encoding schemes, presented in Table~\ref{tab:position_ablation}.
DaPE consistently outperforms all baselines across all tasks. Specifically, DaPE improves zero-shot accuracy by \textbf{+1.8\%}, Recall@1 by \textbf{+2.5\%}, and SLA by a substantial \textbf{+3.0\%} over the strongest baseline (relative encoding). These results demonstrate that incorporating domain-specific spatial priors enhances semantic alignment, generalization, and interpretability in road damage analysis. There are two additional important findings:
1) Removing positional encoding entirely leads to a significant performance drop across all metrics. The zero-shot accuracy drops to 72.1\%, and SLA falls below 54\%, highlighting that explicit spatial priors are critical for accurately understanding and localizing road defects.  These results underscore the sensitivity of Transformer-based models to the absence of spatial signals, particularly in structured domains such as road imagery.  2) The relative position encoding consistently outperforms both sinusoidal and learnable absolute encodings. While learnable absolute encodings show moderate improvements over fixed ones, relative encodings achieve better alignment in tasks requiring spatial awareness—suggesting that modeling relationships between patches, rather than absolute positions alone, better supports semantic grounding.

\begin{table}[!ht]
    \centering
    \caption{Ablation study on the effectiveness of Disease-aware Position Encoding (DaPE). Best results are highlighted in \textbf{bold}, and second-best results are underlined.}
    \resizebox{\linewidth}{!}{
    \begin{tabular}{lcccc}
        \toprule
        \specialrule{0.8pt}{0pt}{0pt}
        \rowcolor[gray]{0.9}
        \textbf{Configuration} & ZS Acc.(\%) & Recall@1(\%) & Recall@5(\%) & SLA(\%) \\
        \specialrule{0.8pt}{0pt}{0pt}
        \midrule
        \textbf{RoadCLIP} w/o Positional Encoding & 72.1 & 51.0 & 71.5 & 53.7 \\
        \textbf{RoadCLIP} w/ Sinusoidal Absolute PE & 74.2 & 53.8 & 72.8 & 56.1 \\
        \textbf{RoadCLIP} w/ Learnable Absolute PE & \underline{75.4} & \underline{54.6} & \underline{74.0} & \underline{57.8} \\
        \textbf{RoadCLIP} w/ Relative PE & 76.8 & 55.9 & 75.3 & 58.9 \\
        \midrule
        \textbf{\textbf{RoadCLIP} w/ DaPE (Proposed)} & \textbf{78.6} & \textbf{58.4} & \textbf{78.1} & \textbf{61.9} \\
        \bottomrule
    \end{tabular}}
    \label{tab:position_ablation}
\end{table}

\begin{table*}[!ht]
    \centering
    \scriptsize
    \caption{Generalization performance comparison of different multimodal models across TD-RD, CNRDD, and CRDDC'22 datasets. Best results are highlighted in \textbf{bold}, second-best are underlined.}
    \resizebox{\textwidth}{!}{
    \begin{tabular}{>{\raggedright\arraybackslash}p{3.2cm}ccc|ccc|ccc}
        \toprule
        \specialrule{0.8pt}{0pt}{0pt}
        \rowcolor[gray]{0.9}
        \multirow{1}{*}{\textbf{Method}} & 
        \multicolumn{3}{c|}{\textbf{TD-RD\cite{xiao2025tdrdtopdownbenchmarkrealtime} (ZS Acc. \%)}} & 
        \multicolumn{3}{c|}{\textbf{CNRDD\cite{zhang2022new} (Recall@1 \%)}} & 
        \multicolumn{3}{c}{\textbf{CRDDC'22\cite{arya2022crowdsensing} (SLA \%)}} \\
        \rowcolor[gray]{0.9}
        & Score & $\Delta$ vs. CLIP & Rank & Score & $\Delta$ vs. CLIP & Rank & Score & $\Delta$ vs. CLIP & Rank \\
        \specialrule{0.8pt}{0pt}{0pt}
        \arrayrulecolor{black}
        CLIP \textcolor{lightgray}{\scriptsize{[ICML21]}}~\cite{radford2021learning} & 65.7 & - & 6 & 38.2 & - & 6 & 42.5 & - & 6 \\
        BLIP-2 \textcolor{lightgray}{\scriptsize{[ICML23]}}~\cite{li2023blip} & 67.4 & +1.7 & 5 & 40.6 & +2.4 & 5 & 44.9 & +2.4 & 5 \\
        LLaVA \textcolor{lightgray}{\scriptsize{[NeurIPS23]}}~\cite{liu2023llava} & 69.1 & +3.4 & 4 & 41.7 & +3.5 & 4 & \underline{47.1} & +4.6 & \underline{2} \\
        GPT-4V \textcolor{lightgray}{\scriptsize{[arXiv23]}}~\cite{openai2023gpt4} & \underline{71.2} & +5.5 & \underline{2} & \underline{44.9} & +6.7 & \underline{2} & 46.8 & +4.3 & 3 \\
        DeepSeek-VL \textcolor{lightgray}{\scriptsize{[arXiv24]}}~\cite{lu2024deepseekvl} & 70.5 & +4.8 & 3 & 43.3 & +5.1 & 3 & 45.2 & +2.7 & 4 \\
        \textbf{\textbf{RoadCLIP} (Ours)} & \textbf{74.3} & +8.6 & \textbf{1} & \textbf{46.9} & +8.7 & \textbf{1} & \textbf{50.1} & +7.6 & \textbf{1} \\
        \bottomrule
    \end{tabular}}
    \label{tab:generalization}
\end{table*}

\subsection{Qualitative Analysis and Visualization}

To assess the interpretability and alignment, we visualize \textbf{RoadCLIP}’s attention using Grad-CAM and cross-modal attention maps to reveal effective grounding of textual descriptions to relevant image regions.

\begin{figure}[!ht]
    \centering
    \includegraphics[width=0.65\linewidth]{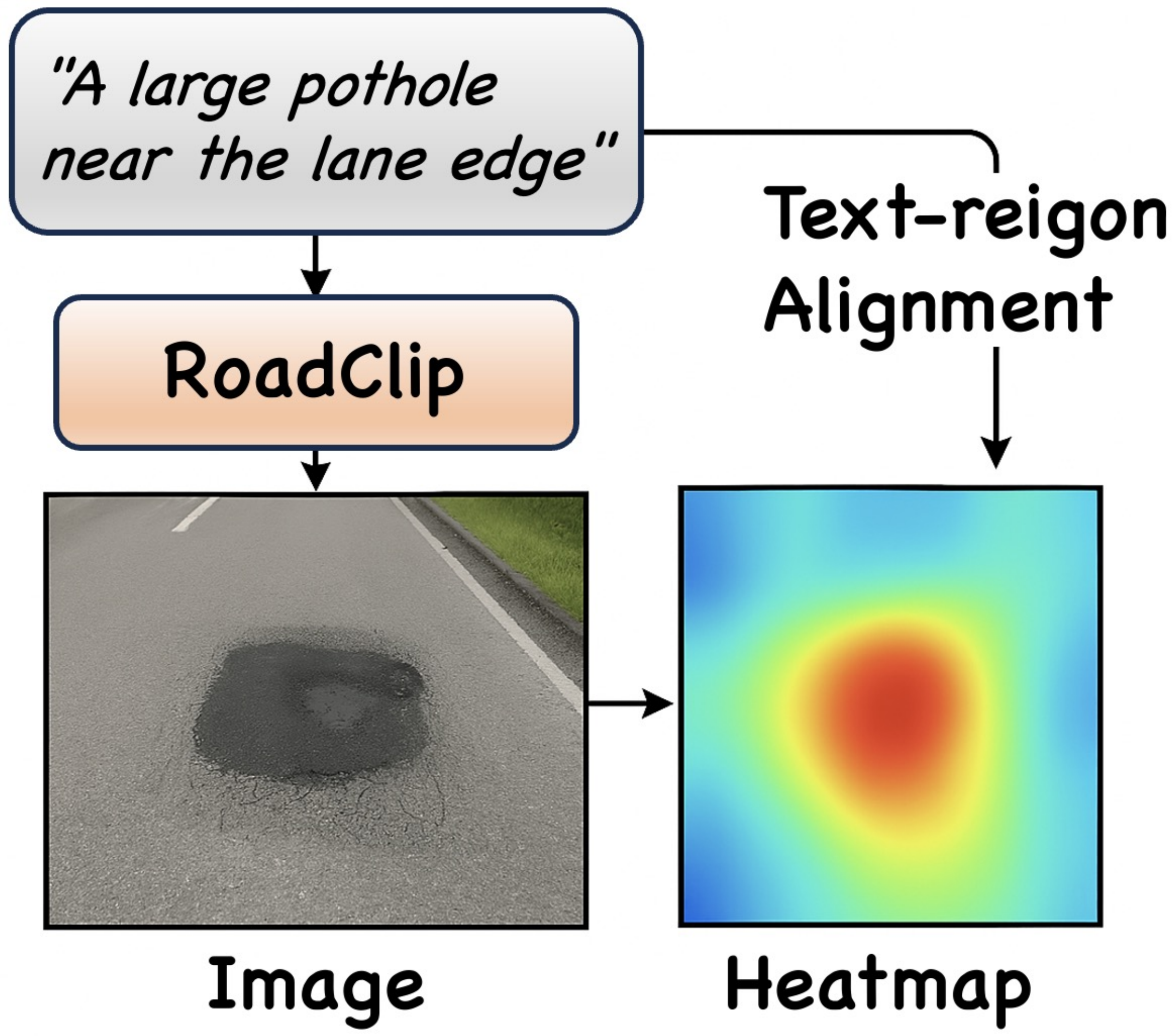}
    \caption{Illustration of the text-region alignment process in \textbf{RoadCLIP}. \textbf{RoadCLIP} encodes text and image, and computes token-wise similarity between text and visual patches. This produces a cross-modal attention map that highlights semantically aligned regions in the image.}
    \label{fig:text_region_alignment}
    \vspace{-0.7em}
\end{figure}

\noindent
\textbf{Fine-Grained Alignment.} 
Figure~\ref{fig:text_region_alignment} illustrates the fine-grained alignment between textual semantics and corresponding image regions enabled by \textbf{RoadCLIP}, through the computation of similarity between textual embeddings and patch-wise image features, resulting in a cross-modal attention map that reflects semantic relevance. Unlike conventional saliency-based visualizations, this approach captures explicit alignment between high-level language semantics and visual content, thereby enabling more interpretable and controllable predictions.

\begin{figure}[!ht]
    \centering
    \includegraphics[width=0.81\linewidth]{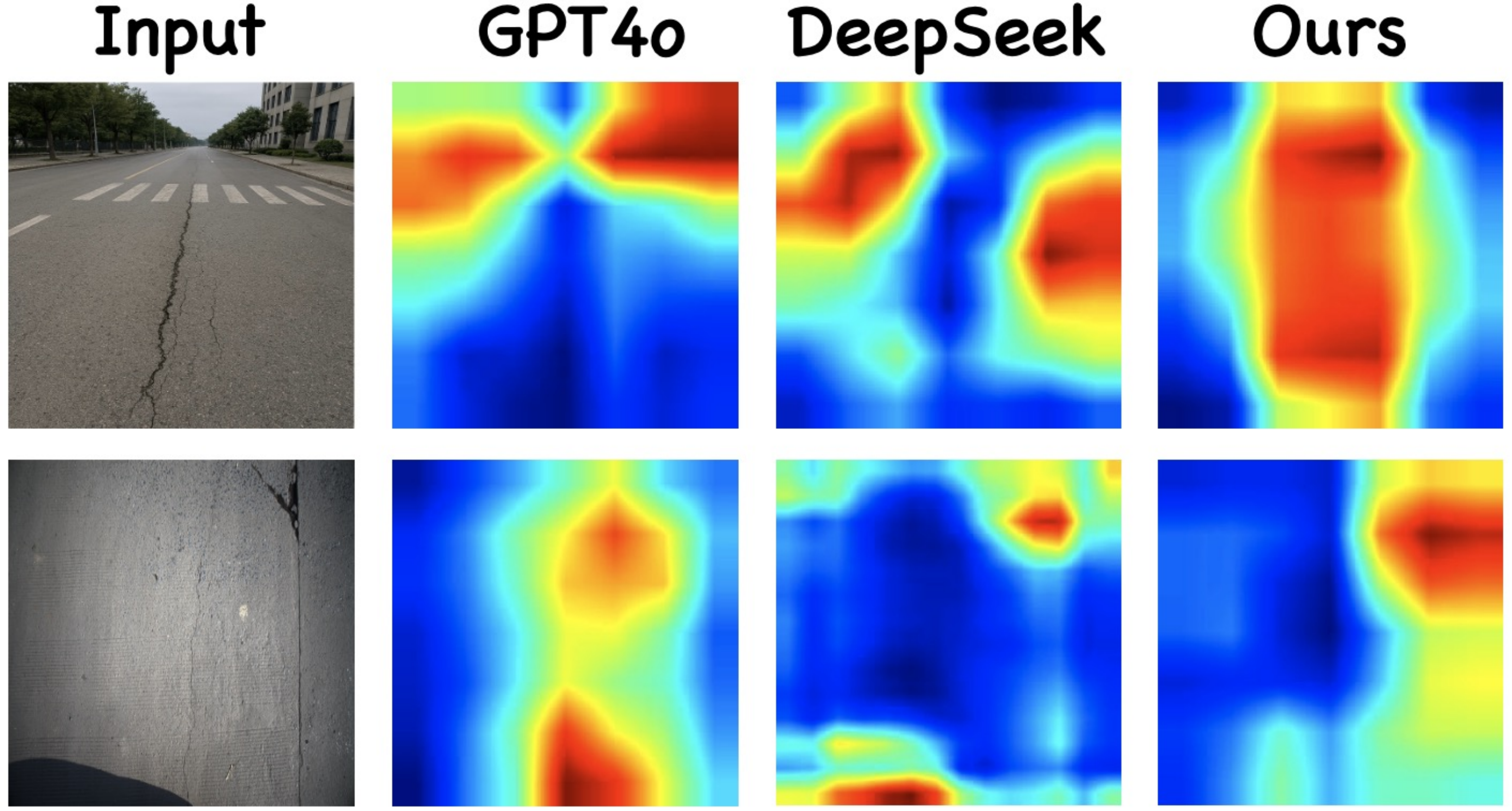}
    \caption{Explainability of \textbf{RoadCLIP} via heatmap++~\cite{chefer2022attentive}.  \textbf{RoadCLIP} demonstrates superior attention localization on road damage regions compared to GPT-4o and DeepSeek-VL.}
    \label{fig:attention_vis}
    \vspace{-1em}
\end{figure}

\noindent
\textbf{Interpretability and Robustness.}  
Figure~\ref{fig:attention_vis} presents comparisons of attention maps generated by \textbf{RoadCLIP} and baseline models. \textbf{RoadCLIP} produces sharper, more localized attention responses on regions indicative of road damage. This indicates that the model captures the structural cues of road damage effectively, benefiting from the integration of DaPE and domain-specific alignment objectives.

\section{Conclusion}
We introduce \textbf{RoadCLIP}, a novel multimodal vision-language framework tailored for fine-grained road damage analysis, along with \textbf{RoadBench}, the first large-scale multimodal benchmark dataset for this domain. Our approach integrates domain-specific knowledge into the CLIP backbone through two key mechanisms: Damage-aware Positional Encoding (DaPE) and Domain-Specific Prior Injection. Experiments across zero-shot classification, image-text retrieval, and semantic localization tasks demonstrate RoadCLIP's substantial improvements over state-of-the-art models, highlighting the value of vision-language integration for infrastructure monitoring. RoadBench and RoadCLIP establish a crucial foundation for future multimodal road damage assessment research.

\section*{Acknowledgment}

This manuscript was co-authored by Oak Ridge National Laboratory (ORNL), operated by UT-Battelle, LLC under Contract No. DE-AC05-00OR22725 with the U.S. Department of Energy. Any subjective views or opinions expressed in this paper do not necessarily represent those of the U.S. Department of Energy or the United States Government.

{\small
\bibliographystyle{ieeenat_fullname}
\bibliography{main}
}

\end{document}